TABLE 1

Numerical Values[a]

| | | | | | | | | | | | | |
|---|---|---|---|---|---|---|---|---|---|---|---|---|
| $(l_{\mathrm{e}}, l_{\mathrm{m}})$ | ... | ... | ... | ... | ... | ... | ... | ... | (10.8, 2) | (20.1, 21) | (52.8, 56) | (55.8, 55) |
| $(l_{e^{-0.5}})$ | ... | ... | ... | ... | ... | ... | ... | ... | ((2), 25) | (14, 29) | (38, 77) | (37, 76) |
| $\sqrt{I(W_l)}$ | ... | ... | ... | ... | ... | ... | ... | ... | 1.62 | 1.02 | 1.02 | 1.15 |
| # | $(\Omega_0, h, \Omega_B h^2)$ | $t_0$ (Gyr) | $S$[b] | $(l_{\mathrm{p1}}, l_{\mathrm{p2}}, l_{\mathrm{p3}})$[d] | $Q_{\mathrm{rms-PS}}^{(\mathrm{no}-Q)}$[e] ($\mu$K) | $AVh^4$[f] ($10^7\mathrm{Mpc}^4$) | $\frac{\delta M}{M}\vert_{8h^{-1}\mathrm{Mpc}}$ | $\beta_I$[h] | FIRS[j] ($\mu$K) | TENERI[l] ($\mu$K) | BARTOL[m] ($\mu$K) | SK94K3[n] ($\mu$K) |
| (1) | (2) | (3) | (4) | (5) | (6) | (7) | (8) | (9) | (10) | (11) | (12) | (13) |
| 1 | (0.1, 0.90, 0.0125) | 13.9 | **0.076** | (217, 522, 810) | 25.6 | 1.89 | 0.59 – 0.73 | **0.19 – 0.24** | 31 – 38 | 29 – 36 | 39 – 48 | 40 – 50 |
| 2 | (0.2, 0.80, 0.0075) | 13.1 | **0.15** | (212, 491, 760) | 23.7 | 0.822 | 1.0 – 1.2 | 0.49 – 0.62 | 30 – 38 | 30 – 37 | 37 – 47 | 38 – 48 |
| 3 | (0.2, 0.75, 0.0125) | 14.0 | **0.13** | (218, 517, 797) | 23.7 | 0.825 | 0.84 – 1.0 | 0.41 – 0.52 | 31 – 38 | 31 – 38 | 39 – 49 | 41 – 51 |
| 4 | (0.2, 0.70, 0.0175) | 15.0 | **0.11** | (226, 549, 845) | 23.7 | 0.828 | 0.68 – 0.85 | 0.34 – 0.42 | 31 – 39 | 31 – 39 | 41 – 52 | 43 – 53 |
| 5 | (0.3, 0.70, 0.0075) | 13.5 | 0.20 | (215, 496, 766) | 22.2 | 0.488 | 1.1 – 1.4 | 0.69 – 0.87 | 30 – 38 | 31 – 38 | 38 – 47 | 39 – 48 |
| 6 | (0.3, 0.65, 0.0125) | 14.5 | **0.17** | (221, 523, 805) | 22.2 | 0.489 | 0.92 – 1.2 | 0.58 – 0.73 | 30 – 38 | 31 – 39 | 40 – 50 | 41 – 51 |
| 7 | (0.3, 0.60, 0.0175) | 15.7 | **0.15** | (231, 558, 857) | 22.2 | 0.489 | 0.75 – 0.93 | 0.47 – 0.59 | 31 – 38 | 32 – 40 | 42 – 52 | 43 – 54 |
| 8 | (0.4, 0.65, 0.0075) | 13.4 | 0.24 | (216, 494, 762) | 21.3 | 0.330 | 1.2 – 1.5 | **0.89 – 1.1** | 30 – 37 | 31 – 39 | 38 – 47 | 39 – 48 |
| 9 | (0.4, 0.60, 0.0125) | 14.5 | 0.21 | (222, 522, 801) | 21.3 | 0.330 | 1.0 – 1.2 | 0.75 – 0.94 | 30 – 38 | 32 – 39 | 40 – 50 | 41 – 51 |
| 10 | (0.4, 0.55, 0.0175) | 15.8 | **0.18** | (232, 557, 853) | 21.3 | 0.330 | 0.82 – 1.0 | 0.61 – 0.76 | 31 – 38 | 32 – 40 | 42 – 52 | 43 – 54 |
| 11 | (0.5, 0.60, 0.0125) | 13.5 | 0.27 | (217, 507, 775) | 20.8 | 0.241 | 1.1 – 1.4 | **0.98 – 1.2** | 30 – 38 | 32 – 39 | 39 – 49 | 40 – 50 |
| 12 | (1.0, 0.50, 0.0125) | 13.0 | **0.45** | (217, 505, 764) | 20.4 | 0.0833 | 1.2 – 1.5 | **1.6 – 2.0** | 30 – 37 | 32 – 40 | 38 – 47 | 39 – 48 |
| Observ. | ... | ... | 0.2 – 0.3[c] | ... | ... | ... | 0.4 – 1.2[g] | 0.25 – 0.80[i] | 23 – 39[k] | 24 – 50[l] | ... | 39 – 69[n] |
| Observ. | ... | ... | ... | ... | ... | ... | ... | ... | 31[k] | 34[l] | ... | 50[n] |

[a] Bandtemperature $\delta T_l$, unless noted otherwise. The ranges on the theoretical predictions are $\pm 1\sigma$ (11% from the DMR normalization), large-scale structure observation ranges are indicative of the current status, and CMB observation ranges are either $\pm 1\sigma$ or $2\sigma$ upper limits. [b] Shape parameter $S = \Omega_0 h \exp[-\Omega_B(\Omega_0 + 1)/\Omega_0]$ (S95). [c] (E.g., Peacock & Dodds 1994). [d] Multipoles of the first three peaks in the $C_l$ spectrum. [e] (BS). [f] Energy-density perturbation power spectrum normalization factor (S95). [g] For a reasonable value of the bias parameter. [h] $\beta_I = 1.3\Omega_0{}^{0.6}\delta M/M(8h^{-1}\mathrm{Mpc})$. [i] (Cole, Fisher, & Weinberg 1995). [j] (Ganga et al. 1994). [k] For an $n = 1$ power-law ($\Omega_0 = 1$) model (not fiducial CDM), with 7% calibration uncertainty (Bond 1995). [l] For an $n = 1$ power-law ($\Omega_0 = 1$) model (not fiducial CDM), with 10% calibration uncertainty (Hancock et al. 1995). [m] (Piccirillo et al. 1995). [n] SK94K$n$ and SK94Q$n$ are the SK94 Ka and Q band $n$-point $W_l$ (Netterfield 1995). Observational numbers are from a reanalysis of the data (Netterfield et al. 1996), and account for 14% calibration uncertainty.

| | | | | | | | | | | | | | |
|---|---|---|---|---|---|---|---|---|---|---|---|---|---|
| (57.2, 64) | (59.4, 55) | (63.6, 54) | (66.2, 73) | (67.0, 69) | (74.4, 95) | (76.5, 75) | (80.6, 75) | (82.3, 59) | (82.9, 73) | (91.7, 73) | (96.0, 95) | (97.6, 109) | (100, 95) |
| (35, 98) | (37, 76) | (37, 76) | (40, 112) | (47, 95) | (28, 118) | (57, 96) | (57, 97) | (39, 84) | (54, 96) | (53, 99) | (76, 116) | (60, 168) | (76, 117) |
| 1.08 | 1.22 | 1.28 | 1.23 | 1.07 | 1.38 | 0.881 | 0.930 | 1.66 | 1.02 | 1.34 | 0.727 | 0.551 | 0.772 |
| SP94Ka[o] | SK94Q3[n] | SK95C3[p] | SP94Q[o] | SK93[q] | PYTHG[r] | SK94K4[n] | SK94Q4[n] | BAM[s] | SK95C4[p] | PYTHL[t] | SK94K5[n] | ARGO[u] | SK94Q5[n] |
| ($\mu$K) | ($\mu$K) | ($\mu$K) | ($\mu$K) | ($\mu$K) | ($\mu$K) | ($\mu$K) | ($\mu$K) | ($\mu$K) | ($\mu$K) | ($\mu$K) | ($\mu$K) | ($\mu$K) | ($\mu$K) |
| (14) | (15) | (16) | (17) | (18) | (19) | (20) | (21) | (22) | (23) | (24) | (25) | (26) | (27) |
| 41 – 51 | 41 – 52 | 42 – 53 | 44 – 55 | 44 – 54 | 46 – 57 | 46 – 58 | 48 – 59 | 48 – 59 | 48 – 60 | 50 – 63 | 52 – 65 | **53 – 66** | 53 – 67 |
| 39 – 49 | 39 – 49 | 40 – 50 | 41 – 52 | 41 – 51 | 43 – 54 | 43 – 54 | 44 – 55 | 45 – 56 | 45 – 56 | 47 – 58 | 48 – 60 | **49 – 61** | 49 – 62 |
| 41 – 52 | 42 – 52 | 43 – 53 | 44 – 55 | 44 – 55 | 46 – 57 | 46 – 58 | 48 – 59 | 48 – 60 | 48 – 60 | 50 – 63 | 52 – 65 | **53 – 66** | 53 – 66 |
| 43 – 54 | 44 – 55 | 45 – 56 | 47 – 58 | 46 – 58 | 49 – 61 | **49 – 61** | 51 – 63 | 51 – 64 | 51 – 64 | 54 – 67 | 55 – 69 | **56 – 70** | 57 – 71 |
| 39 – 49 | 40 – 50 | 41 – 51 | 42 – 52 | 41 – 51 | 43 – 54 | 43 – 54 | 44 – 55 | 45 – 56 | 45 – 56 | 47 – 58 | 48 – 60 | **49 – 61** | 49 – 61 |
| 42 – 52 | 42 – 53 | 43 – 54 | 44 – 55 | 44 – 55 | 46 – 58 | **47 – 58** | 48 – 60 | 48 – 60 | 48 – 60 | 50 – 63 | 52 – 65 | **53 – 66** | 53 – 66 |
| 44 – 55 | 44 – 55 | 46 – 57 | 47 – 59 | 47 – 58 | 50 – 62 | **50 – 62** | 51 – 64 | 52 – 64 | 51 – 64 | 54 – 67 | 56 – 69 | **57 – 71** | 57 – 71 |
| 39 – 49 | 39 – 49 | 40 – 50 | 41 – 51 | 41 – 51 | 43 – 54 | 43 – 53 | 44 – 55 | 44 – 55 | 44 – 55 | 46 – 58 | 47 – 59 | **48 – 60** | 48 – 60 |
| 42 – 52 | 42 – 52 | 43 – 54 | 44 – 55 | 44 – 55 | 46 – 58 | 46 – 57 | 47 – 59 | 48 – 60 | 48 – 60 | 50 – 62 | 51 – 64 | **52 – 65** | 53 – 66 |
| 44 – 55 | 44 – 55 | 46 – 57 | 47 – 58 | 46 – 58 | 49 – 61 | **49 – 61** | 51 – 63 | 51 – 64 | 51 – 64 | 54 – 67 | 55 – 69 | **56 – 70** | 57 – 71 |
| 41 – 51 | 41 – 51 | 42 – 52 | 43 – 54 | 43 – 53 | 45 – 56 | 45 – 56 | 46 – 57 | 46 – 58 | 46 – 58 | 48 – 60 | 49 – 62 | **51 – 63** | 51 – 63 |
| 39 – 49 | 39 – 49 | 40 – 50 | 41 – 51 | 41 – 51 | 43 – 53 | 42 – 53 | 43 – 54 | 44 – 55 | 44 – 54 | 45 – 57 | 46 – 58 | **47 – 59** | 47 – 59 |
| 22 – 44[o] | 27 – 62[n] | ... | 31 – 57[o] | 26 – 49[q] | 49 – 85[r] | 24 – 46[n] | $< 76$[n] | ... | ... | 42 – 69[t] | 33 – 62[n] | 34 – 45[u] | $< 59$[n] |
| 29[o] | 41[n] | ... | 40[o] | 36[q] | 64[r] | 32[n] | 12[n] | ... | ... | 54[t] | 44[n] | 39[u] | 17[n] |

[o]Data-weighted $W_l$ (Gundersen et al. 1995), with 15% calibration uncertainty. [p]SK95C$n$ and SK95R$n$ are the SK95 Q band NCP cap and ring $n$-point chops (Netterfield et al. 1996). [q](Netterfield 1995), with 15% calibration uncertainty. [r]PYTHG,I are the PYTHON additive large- and small-chop $W_l$ (Alvarez et al. 1996). Calibration uncertainty is 20%. [s](Halpern et al. 1993). [t]PYTHL,S are the PYTHON subtractive large- and small-chop $W_l$ (Platt et al. 1995), with three years of data for L, and one year for S. Calibration uncertainty is 20%. [u](de Bernardis et al. 1994), fiducial CDM, with 5% calibration uncertainty.

| | | | | | | | | | | | | | | |
|---|---|---|---|---|---|---|---|---|---|---|---|---|---|---|
| (108, 94) | (114, 127) | (115, 115) | (120, 115) | (125, 123) | (133, 145) | (134, 134) | (135, 117) | (139, 134) | (139, 150) | (142, 155) | (144, 157) | (152, 153) | (157, 140) | (157, 154) |
| (76, 115) | (70, 196) | (95, 136) | (95, 136) | (84, 168) | (80, 224) | (115, 155) | (97, 141) | (115, 156) | (83, 232) | (85, 240) | (86, 243) | (134, 175) | (119, 165) | (134, 176) |
| 0.810 | 1.41 | 0.620 | 0.665 | 0.895 | 1.51 | 0.542 | 0.755 | 0.586 | 1.55 | 1.49 | 1.51 | 0.481 | 0.667 | 0.524 |
| SK95C5[p] ($\mu$K) (28) | MAX4 6,9[v] ($\mu$K) (29) | SK94K6[n] ($\mu$K) (30) | SK94Q6[n] ($\mu$K) (31) | PYTHI[r] ($\mu$K) (32) | MAX4 3.5[w] ($\mu$K) (33) | SK94K7[n] ($\mu$K) (34) | SK95C6[p] ($\mu$K) (35) | SK94Q7[n] ($\mu$K) (36) | MAX5[x] ($\mu$K) (37) | MAX3 G[y] ($\mu$K) (38) | MAX3 M[z] ($\mu$K) (39) | SK94K8[p] ($\mu$K) (40) | SK95C7[p] ($\mu$K) (41) | SK94Q8[p] ($\mu$K) (42) |
| 55 – 68 | 56 – 70 | **58 – 72** | 59 – 74 | 59 – 73 | 59 – 74 | 63 – 78 | 61 – 76 | 64 – 79 | 59 – 74 | 60 – 75 | **60 – 75** | 67 – 84 | 66 – 82 | 68 – 85 |
| 50 – 63 | 52 – 65 | **53 – 66** | 54 – 68 | 54 – 68 | 55 – 68 | 58 – 72 | 56 – 70 | 59 – 73 | 55 – 69 | 56 – 69 | **56 – 69** | 62 – 77 | 61 – 76 | 63 – 78 |
| 54 – 68 | 56 – 70 | **57 – 72** | 59 – 73 | 58 – 73 | 59 – 74 | 63 – 78 | 61 – 76 | 63 – 79 | 60 – 74 | 60 – 75 | **60 – 75** | 67 – 84 | 66 – 82 | 68 – 85 |
| 58 – 73 | 60 – 75 | **61 – 77** | 63 – 78 | 63 – 78 | 64 – 79 | 67 – 84 | 65 – 82 | 68 – 85 | 64 – 80 | 65 – 81 | **65 – 81** | 72 – 90 | 71 – 89 | 73 – 91 |
| 50 – 63 | 52 – 65 | **53 – 66** | 54 – 67 | 54 – 67 | 55 – 68 | 58 – 72 | 56 – 70 | 58 – 73 | 55 – 69 | 56 – 69 | **56 – 70** | 62 – 77 | 61 – 76 | 63 – 78 |
| 54 – 68 | 57 – 71 | **57 – 72** | 59 – 73 | 59 – 73 | 59 – 74 | 63 – 78 | 61 – 76 | 63 – 79 | 60 – 75 | 60 – 75 | **61 – 76** | 67 – 84 | 66 – 82 | 68 – 85 |
| 59 – 73 | 61 – 76 | **62 – 77** | 63 – 79 | 63 – 79 | 64 – 80 | 67 – 84 | 66 – 82 | 68 – 85 | 65 – 81 | 66 – 82 | **66 – 82** | 72 – 90 | 72 – 89 | 73 – 92 |
| 50 – 62 | 51 – 64 | **52 – 65** | 53 – 66 | 53 – 66 | 54 – 67 | 57 – 71 | 55 – 69 | 57 – 72 | 54 – 68 | 55 – 68 | **55 – 68** | 61 – 76 | 60 – 74 | 61 – 77 |
| 54 – 67 | 56 – 70 | **57 – 71** | 58 – 72 | 58 – 72 | 59 – 73 | 62 – 77 | 60 – 75 | 63 – 78 | 59 – 74 | 60 – 74 | **60 – 75** | 66 – 83 | 65 – 81 | 67 – 84 |
| 58 – 72 | 60 – 75 | **61 – 76** | 62 – 78 | 63 – 78 | 64 – 79 | 67 – 83 | 65 – 81 | 68 – 85 | 64 – 80 | 65 – 81 | **65 – 81** | 72 – 89 | 71 – 88 | 73 – 91 |
| 52 – 64 | 54 – 67 | **54 – 68** | 55 – 69 | 56 – 69 | 56 – 70 | 59 – 74 | 58 – 72 | 60 – 75 | 57 – 71 | 57 – 71 | **57 – 71** | 63 – 79 | 62 – 78 | 64 – 80 |
| 48 – 60 | 50 – 63 | 50 – 63 | 51 – 64 | 52 – 64 | 52 – 65 | 55 – 68 | 53 – 67 | 55 – 69 | 53 – 66 | 53 – 66 | **53 – 66** | 59 – 73 | 58 – 72 | 59 – 74 |
| … | 29 – 68[v] | 22 – 50[n] | $< 63$[n] | … | 42 – 87[w] | $< 76$[n] | … | $< 120$[n] | 25 – 71[x] | 60 – 96[y] | 16 – 36[z] | $< 130$[p] | … | $< 120$[p] |
| … | … | 33[n] | 25[n] | … | … | 35[n] | … | 67[n] | … | 74[y] | 23[z] | 66[p] | … | 64[p] |

[v] MAX4 6 and 9 cm$^{-1}$ $W_l$ (Devlin et al. 1994; Clapp et al. 1994; as reanalyzed by Tanaka et al. 1995, hereafter T95; individual channel numbers are courtesy of S. Tanaka, private communication 1995). The observational range is a composite of 3 sets of 6 and 9 cm$^{-1}$ measurements — $42^{+18}_{-13}\mu$K, $54^{+25}_{-17}\mu$K (GUM); $(15, < 72)\mu$K, $42^{+26}_{-18}\mu$K (SH); $51^{+29}_{-21}\mu$K, $41^{+27}_{-19}\mu$K (ID) — estimated by discarding the two largest upper limits and the two smallest lower limits (among the five detections). Observational error bars account for calibration uncertainty of 10%. [w] MAX4 3.5 cm$^{-1}$ $W_l$. See previous footnote. The observational range is a composite of 3 sets of 3.5 cm$^{-1}$ measurements — $79^{+29}_{-23}\mu$K (GUM); $61^{+26}_{-19}\mu$K (SH); $56^{+26}_{-18}\mu$K (ID) — estimated by discarding the largest upper limit and the smallest lower limit. [x] MAX5 (T95). The observational range is a composite of two measurements — $33^{+11}_{-8}\mu$K, $52^{+19}_{-11}\mu$K (HR5127, PH; T95) — estimated using the largest upper limit and the smallest lower limit. Calibration uncertainty is 10%. A third MAX5 scan is under analysis (Lim et al. 1996). [y] MAX3 GUM (Gundersen et al. 1993, as recomputed by J. Gundersen, private communication 1995). All MAX GUM $W_l$ are for a single scan and ignore sky rotation. Calibration uncertainty is 10%. [z] MAX3 MUP (Meinhold et al. 1993, as recomputed by J. Gundersen, private communication 1995). Note that dust has been subtracted.

| | | | | | | | | | | | | | | |
|---|---|---|---|---|---|---|---|---|---|---|---|---|---|---|
| (159, 151) | (170, 171) | (170, 173) | (176, 173) | (177, 160) | (178, 183) | (197, 180) | (217, 199) | (234, 237) | (237, 220) | (257, 241) | (263, 270) | (277, 262) | (286, 302) | (297, 283) |
| (83, 234) | (115, 236) | (153, 194) | (153, 195) | (140, 184) | (133, 239) | (159, 204) | (178, 222) | (182, 301) | (199, 244) | (221, 265) | (181, 375) | (241, 286) | (247, 365) | (263, 307) |
| 1.45 | 1.13 | 0.432 | 0.474 | 0.594 | 0.684 | 0.543 | 0.496 | 0.852 | 0.461 | 0.425 | 1.14 | 0.398 | 0.702 | 0.371 |
| MSAM2[aa] ($\mu$K) (43) | SK95R3[p] ($\mu$K) (44) | SK94K9[p] ($\mu$K) (45) | SK94Q9[p] ($\mu$K) (46) | SK95C8[p] ($\mu$K) (47) | PYTHS[t] ($\mu$K) (48) | SK95C9[p] ($\mu$K) (49) | SK95C10[p] ($\mu$K) (50) | SK95R4[p] ($\mu$K) (51) | SK95C11[p] ($\mu$K) (52) | SK95C12[p] ($\mu$K) (53) | MSAM3[bb] ($\mu$K) (54) | SK95C13[p] ($\mu$K) (55) | SK95R5[p] ($\mu$K) (56) | SK95C14[p] ($\mu$K) (57) |
| **59 – 74** | 65 – 82 | 71 – 88 | **71 – 89** | 69 – 86 | 68 – 85 | 72 – 89 | 73 – 91 | 69 – 86 | 73 – 91 | 71 – 89 | **65 – 81** | 69 – 85 | 63 – 78 | 65 – 81 |
| **55 – 69** | 60 – 75 | 65 – 81 | **66 – 82** | 64 – 80 | 63 – 79 | 66 – 82 | 67 – 83 | 63 – 79 | 67 – 83 | 65 – 81 | **60 – 75** | 62 – 78 | 58 – 72 | 59 – 74 |
| **59 – 74** | 65 – 82 | 71 – 88 | **71 – 89** | 69 – 86 | 68 – 85 | 72 – 90 | 73 – 91 | 69 – 86 | 73 – 91 | 72 – 89 | **65 – 81** | 69 – 86 | 63 – 78 | 65 – 81 |
| **64 – 80** | 71 – 88 | 76 – 95 | 77 – 96 | 75 – 93 | **74 – 93** | 78 – 97 | 80 – 99 | 76 – 95 | 80 – 100 | 79 – 99 | **71 – 89** | 77 – 96 | 70 – 87 | 73 – 91 |
| **55 – 69** | 60 – 75 | 65 – 81 | **66 – 82** | 64 – 80 | 63 – 79 | 66 – 83 | 67 – 84 | 64 – 79 | 67 – 84 | 66 – 82 | **61 – 76** | 63 – 79 | 58 – 73 | 60 – 75 |
| **60 – 75** | 66 – 82 | 71 – 88 | **72 – 89** | 70 – 87 | 69 – 86 | 72 – 90 | 74 – 92 | 70 – 87 | 74 – 92 | 73 – 91 | **66 – 83** | 70 – 88 | 64 – 80 | 67 – 83 |
| **65 – 81** | 72 – 89 | 77 – 96 | 78 – 97 | 76 – 94 | **75 – 94** | 79 – 98 | 81 – 100 | 77 – 96 | 82 – 100 | 81 – 100 | **73 – 91** | 79 – 99 | 72 – 90 | 76 – 94 |
| **54 – 68** | 59 – 74 | 64 – 80 | **65 – 81** | 63 – 78 | 62 – 77 | 65 – 81 | 66 – 83 | 63 – 78 | 66 – 83 | 65 – 81 | **60 – 75** | 62 – 78 | 58 – 72 | 59 – 74 |
| **59 – 74** | 65 – 81 | 70 – 87 | **71 – 88** | 69 – 86 | 68 – 85 | 71 – 89 | 73 – 91 | 69 – 86 | 73 – 91 | 72 – 90 | **66 – 82** | 69 – 87 | 63 – 79 | 66 – 82 |
| **64 – 80** | 71 – 89 | 76 – 95 | 77 – 96 | 75 – 93 | **74 – 93** | 78 – 98 | 80 – 100 | 77 – 96 | 81 – 100 | 81 – 100 | **72 – 90** | 79 – 98 | 71 – 89 | 75 – 94 |
| **56 – 70** | 62 – 77 | 67 – 83 | **67 – 84** | 66 – 82 | 65 – 81 | 68 – 85 | 69 – 86 | 65 – 82 | 69 – 86 | 68 – 84 | **62 – 77** | 65 – 81 | 60 – 74 | 61 – 77 |
| **53 – 66** | 57 – 72 | 62 – 77 | **62 – 78** | 61 – 76 | 60 – 75 | 63 – 78 | 64 – 80 | 61 – 76 | 64 – 80 | 63 – 78 | **58 – 72** | 60 – 75 | 55 – 69 | 57 – 71 |
| 24 – 44[aa] | ... | $<$ 150[p] | 90 – 200[p] | ... | 46 – 73[t] | ... | ... | ... | ... | ... | 30 – 52[bb] | ... | ... | ... |
| 33[aa] | ... | 5.8[p] | 140[p] | ... | 59[t] | ... | ... | ... | ... | ... | 40[bb] | ... | ... | ... |

[aa]Second year MSAM 2-beam $W_l$ (Cheng et al. 1995, hereafter C95). Unless given in C95, MSAM central values (cv) are the geometric mean of the upper and lower $2\sigma$ values, and the $1\sigma$ values are $= \mathrm{cv}(2\sigma/\mathrm{cv})^{1/1.7}$ (D. Cottingham, private communication 1995), with 10% calibration uncertainty. The first year measurement gives $53^{+15}_{-12}\mu$K (C95). [bb]Second year MSAM 3-beam $W_l$ (C95). See previous footnote. The first year measurement gives $53^{+12}_{-11}\mu$K (C95).

| | | | | | | | | | | | | | | |
|---|---|---|---|---|---|---|---|---|---|---|---|---|---|---|
| (316, 302) | (329, 382) | (332, 369) | (333, 322) | (357, 346) | (382, 371) | (404, 394) | (477, 539) | (559, 608) | (598, 537) | (759, 804) | (2020, 1700) | (2360, 2410) | (3250, 3670) | (4010, 4210) |
| (282, 326) | (213, 570) | (315, 431) | (301, 345) | (325, 369) | (352, 393) | (374, 416) | (297, 825) | (412, 815) | (361, 754) | (591, 1030) | (1140, 2390) | (1340, 3680) | (2020, 5600) | (2830, 5770) |
| 0.352 | 1.56 | 0.585 | 0.334 | 0.312 | 0.288 | 0.274 | 1.18 | 0.781 | 1.41 | 0.415 | 1.50 | 1.73 | 1.27 | 0.933 |
| SK95C15[p] ($\mu$K) (58) | WDH1[cc] ($\mu$K) (59) | SK95R6[p] ($\mu$K) (60) | SK95C16[p] ($\mu$K) (61) | SK95C17[p] ($\mu$K) (62) | SK95C18[p] ($\mu$K) (63) | SK95C19[p] ($\mu$K) (64) | WDI[dd] ($\mu$K) (65) | WDH2[cc] ($\mu$K) (66) | OVROL[ee] ($\mu$K) (67) | WDH3[cc] ($\mu$K) (68) | OVROS[ee] ($\mu$K) (69) | SUZIL2[ff] ($\mu$K) (70) | SUZIS2[ff] ($\mu$K) (71) | SUZIS3[ff] ($\mu$K) (72) |
| 61 – 76 | 60 – 74 | 55 – 68 | 57 – 71 | 53 – 66 | 50 – 62 | 49 – 61 | **55 – 69** | 52 – 65 | 51 – 64 | 46 – 57 | 24 – 29 | 23 – 29 | 17 – 21 | 7.0 – 8.7 |
| 56 – 70 | 56 – 70 | 53 – 66 | 53 – 66 | 50 – 63 | 50 – 62 | 50 – 63 | **52 – 65** | 50 – 63 | 50 – 62 | 44 – 55 | 21 – 26 | 21 – 27 | 15 – 19 | 5.6 – 7.0 |
| 61 – 77 | 60 – 75 | 55 – 69 | 58 – 72 | 54 – 67 | 51 – 64 | 51 – 63 | **56 – 70** | 54 – 67 | 53 – 66 | 48 – 60 | 25 – 31 | 25 – 31 | 18 – 22 | 7.5 – 9.3 |
| 69 – 87 | 65 – 81 | 60 – 75 | 65 – 81 | 59 – 74 | 55 – 68 | 52 – 65 | **60 – 75** | 56 – 70 | 56 – 70 | 51 – 63 | 28 – 35 | 27 – 34 | 20 – 24 | 8.9 – 11 |
| 57 – 71 | 57 – 71 | 53 – 66 | 54 – 67 | 51 – 64 | 50 – 63 | 51 – 64 | **53 – 67** | 51 – 64 | 51 – 63 | 45 – 56 | 22 – 27 | 22 – 27 | 16 – 19 | 5.9 – 7.3 |
| 63 – 79 | 61 – 76 | 57 – 71 | 59 – 74 | 55 – 69 | 52 – 65 | 51 – 64 | **57 – 72** | 55 – 69 | 54 – 68 | 50 – 62 | 26 – 33 | 26 – 32 | 18 – 23 | 7.9 – 9.8 |
| 72 – 90 | 66 – 83 | 62 – 77 | 68 – 84 | 62 – 77 | 57 – 71 | 54 – 67 | **62 – 77** | 58 – 72 | 58 – 72 | 52 – 65 | 29 – 37 | 28 – 35 | 20 – 26 | 9.5 – 12 |
| 56 – 70 | 56 – 70 | 53 – 66 | 53 – 67 | 51 – 63 | 50 – 63 | 51 – 64 | **53 – 66** | 51 – 64 | 50 – 63 | 45 – 56 | 22 – 27 | 22 – 27 | 16 – 19 | 5.9 – 7.4 |
| 62 – 78 | 61 – 76 | 56 – 70 | 59 – 73 | 55 – 68 | 52 – 65 | 51 – 64 | **57 – 71** | 55 – 69 | 54 – 68 | 50 – 62 | 26 – 33 | 26 – 32 | 18 – 23 | 8.0 – 9.9 |
| 71 – 89 | 66 – 82 | 61 – 77 | 67 – 84 | 62 – 77 | 57 – 71 | 53 – 67 | **61 – 77** | 58 – 73 | 58 – 72 | 53 – 66 | 30 – 37 | 29 – 36 | 21 – 26 | 9.7 – 12 |
| 58 – 72 | 58 – 72 | 53 – 67 | 55 – 68 | 51 – 64 | 50 – 62 | 50 – 62 | **55 – 68** | 53 – 66 | 52 – 65 | 49 – 61 | 25 – 32 | 25 – 31 | 18 – 22 | 7.5 – 9.4 |
| 54 – 67 | 54 – 67 | 50 – 62 | 51 – 63 | 48 – 60 | 47 – 58 | 47 – 59 | **52 – 65** | 51 – 63 | 50 – 62 | 47 – 58 | 25 – 31 | 24 – 30 | 17 – 21 | 7.3 – 9.1 |
| ... | 43 – 120[cc] | ... | ... | ... | ... | ... | $<$ 50[dd] | $<$ 88[cc] | ... | $<$ 160[cc] | ... | ... | ... | ... |
| ... | 77[cc] | ... | ... | ... | ... | ... | 0[dd] | 30[cc] | ... | 60[cc] | ... | ... | ... | ... |

[cc] WDH$n$ are the WD $n^{\rm th}$-harmonic $W_l$ (Griffin et al. 1995), with calibration uncertainty of 30%. [dd] (Tucker et al. 1993). Calibration uncertainty is accounted for by multiplying the $2\sigma$ upper limit by $\sqrt{1.3}$ (J. Peterson, private communication 1995). [ee] (Myers, Leitch, & Readhead 1996). [ff] SUZIL2, S2, and S3 are the SuZIE large 2-beam, and small 2- and 3-beam $W_l$ (Church et al. 1996; Ganga et al. 1996).

# CMB ANISOTROPY IN COBE-DMR-NORMALIZED FLAT $\Lambda$ CDM COSMOGONY

Bharat Ratra[1], and Naoshi Sugiyama[2]
[1]*Joseph Henry Laboratories, Princeton University, Princeton, NJ 08544*
[2]*Department of Physics and Research Center for the Early Universe, University of Tokyo, Tokyo 113, Japan*

## ABSTRACT

We compute the cosmic microwave background (CMB) anisotropy in a low-density, flat, cosmological constant, cold dark matter model which is normalized to the two-year COBE DMR sky map. Although conclusions regarding model viability must remain tentative until systematic effects are better understood, there are mild indications that these models have more intermediate scale power than is indicated by presently available CMB anisotropy observational data, with old ($t_0 \gtrsim 15 - 16$Gyr), high baryon density ($\Omega_B \gtrsim 0.0175h^{-2}$), low density ($\Omega_0 \sim 0.2 - 0.4$) models doing the worst.

## 1. INTRODUCTION

Recent determinations of the Hubble parameter $h = H_0/(100\mathrm{km\,s^{-1}Mpc^{-1}})$ and the age of the universe $t_0$, combined with other observational evidence, suggests that the mass density parameter $\Omega_0$ is small. (We emphasize, however, that if all the mass were baryonic, there are weak indications that some of it must be in a form that does not take part in nucleosynthesis, e.g., Reeves 1994; Kernan & Krauss 1994.) A low-density cold dark matter (CDM) model with flat spatial sections and a cosmological constant $\Lambda$, with an initial epoch of inflation, might eventually prove to be a reasonable model of the universe (Peebles 1988)[4]. Normalized to the two-year DMR sky maps (Bunn & Sugiyama 1995, hereafter BS; Sugiyama 1995, hereafter S95; Stompor, Górski, & Banday 1995, hereafter SGB), the model is mostly consistent with large-scale structure observations when $\Omega_0 \sim 0.3$, although there are mild indications that the model has an excess of large-scale structure power on intermediate scales (SGB; Scott, Silk, & White 1995, hereafter SSW). Here we examine the compatability of the primary[5] CMB anisotropy predictions of this DMR-normalized inflation model with what has been measured on angular scales smaller than that probed by the DMR.

## 2. SUMMARY OF COMPUTATION

The parameters characterizing models 1 – 11 (Table 1, col. [2]) are chosen to be roughly consistent with small-scale dynamical estimates of $\Omega_0$, measurements of $h$ and $t_0$, and nucleosynthesis bounds on $\Omega_B h^2$. Model 12 is the fiducial flat CDM model, and is only included for the purpose of a comparison. To make the problem tractable, effects of reionization, tilt, and gravity waves are ignored here.

---

[4]The original motivation for flat spatial sections no longer holds — there are open inflation models (Ratra & Peebles 1994; Górski et al. 1995; Bucher & Turok 1995; Linde 1995; Amendola, Baccigalupi, & Occhionero 1995; Liddle et al. 1995; Yamamoto & Bunn 1995; Ratra et al. 1995). Also, quantum-mechanical radiative corrections affect $\Lambda$ at all energy scales; what happens to space curvature is less clear, but on energy scales below the Planck scale it seems to be consistent to ignore the effects of quantum mechanics on space curvature.

[5]For this class of models, secondary CMB anisotropies are likely to be insignificant on all but the smallest scales we consider here.

The computation follows Ratra et al. (1995, hereafter RBGS), and the details may be found there. To normalize the model we use the quadrupole-excluded (ecliptic coordinates) DMR normalization of BS (col. [6]). We follow SGB and also account for systematic uncertainties in the DMR normalization by assigning total ($1\sigma$) error bars of $\pm 11\%$ to the $Q_{\rm rms-PS}^{\rm (no-Q)}$ normalization. These systematics are: the difference between the ecliptic and galactic coordinate maps; the effect of including or excluding the quadrupole in the analysis; the effects of varying $\Omega_B$ and $h$ on the DMR-scale CMB anisotropy (BS normalized at fixed $\Omega_B$ and $h$); and, the numerical uncertainty in the CMB anisotropy computation. With $\delta T/T = \sum_{l,m} a_{lm} Y_{lm}$, the rms temperature anisotropy, seen through a window $W_l$, is $(\delta T/T)_{\rm rms} = \left(\sum_{l=2}^{\infty}(2l+1)C_l W_l/(4\pi)\right)^{1/2}$, where $C_l = \langle |a_{lm}|^2 \rangle$. Defining $I(W_l) = \sum_{l=2}^{\infty}(l+0.5)W_l/[l(l+1)]$, the bandtemperature $\delta T_l = \delta T_{\rm rms}/\sqrt{I(W_l)}$, and the effective multipole $l_{\rm e} = I(lW_l)/I(W_l)$ (Bond 1995). Two CMB anisotropy bandtemperature spectra, at the extremes of the range of models we consider, are illustrated in the figures. The first line of the table gives $l_{\rm e}$ (which is where the predictions and data are placed in the figures; $l_{\rm e}$ is mostly just a convenient measure for ordering the $W_l$), and $l_{\rm m}$, the multipole where $W_l$ peaks. The second line gives $l_{e^{-0.5}}$, the two multipoles at which $W_{l_{e^{-0.5}}} = e^{-0.5} W_{l_{\rm m}}$ (except for FIRS). The third line is the conversion factor between $\delta T_l$ and $\delta T_{\rm rms}$. The DMR-normalized $1\sigma$ range of the bandtemperature predictions for models 1 – 12 are tabulated for the experimental $W_l$ in columns (10) – (72) of the table.

As discussed in RBGS, the observational data points have been converted to bandtemperature, $\delta T_l$, with $1\sigma$ error bars for those points where there is a $2\sigma$ detection away from 0 or are $2\sigma$ upper limits when there is not a $2\sigma$ detection away from 0. These error bars account for the size of the observational sample, as well as the uncertainty introduced due to there being only one observable universe. Absolute calibration uncertainty has been added in quadrature to the error bars. It is not strictly correct to account for calibration uncertainty in this manner, but given the large uncertainty we have adopted this procedure. The last line of the table gives the central value, and the penultimate line the range, of the observational data. Due caution must be exercised when comparing different observational results, as well as when comparing them to model predictions (RBGS).

The observational estimates are based on an assumption of the form of the CMB anisotropy spectrum — usually either a flat or power-law bandpower or a gaussian correlation function. While none of these forms are a good approximation to the spectra we consider here, it is clear that one can, at this point (given the size of the other uncertainties), ignore the error due to this approximation for narrow windows like SK94 or SK95 (provided the CMB spectrum is not changing rapidly through the window). It likely that this assumption cannot be justified for broad windows, like SuZIE or FIRS, and it would be useful to study such cases using more realistic CMB spectra, as has been done for the DMR (e.g., BS; S95; Górski et al. 1995; SGB).

## 3. DISCUSSION

The boldface entries in cols. (4), (8), and (9) of the table are disfavoured by what are thought to be reasonable large-scale structure estimates. The allowed parameter space can be further constrained by using other observational data. For instance, estimates of the baryon mass fraction of clusters (White et al. 1993) are difficult to reconcile with models 5, 8, 11, and 12. Cluster abundances (White, Efstathiou, & Frenk 1993) disfavour the $\delta M/M(8h^{-1}\mathrm{Mpc})$ values of models 1, 3, 4, 11, and 12, and put pressure on models 7 and 8.

Models 4, 7, and 10 would seem to be difficult to reconcile with CMB anisotropy data, and as indicated by the boldface entries in the CMB section of the table, there are mild indications that flat $\Lambda$ models have an excess of intermediate-scale CMB anisotropy power. (It is interesting that lowering $\Omega_B$ or raising $h$ helps resolves this problem, but this would exacerbate the excess intermediate-scale structure power problem, SGB; SSW) However, better control of systematic uncertainties (especially calibration uncertainty) will be needed before it will be possible to draw robust conclusions about the viability of these theoretical models. We emphasize that present CMB observational data does not strongly discriminate between models.

A comparison between models and CMB data like that done here has discriminative power for narrow windows like those of SK (as is indicated by the larger spread in the predictions for models in, e.g., col. [40]). Such a comparison is not as discriminative for wider windows (e.g., OVROL, col. [67]) since they average over a large range in $l$, and in cases like this a direct maximum likelihood type comparison will be needed to fully utilize the discriminative power of the data.

If it turns out that flat $\Lambda$ CDM models cannot be reconciled with observational data, one way out would be to enlarge parameter space by including tilt, gravity waves, and/or reionization (SSW; Ostriker & Steinhardt 1995). These parameters might soon be constrained by the OVRO and SuZIE CMB anisotropy observations. Another possibility is to consider a time-variable $\Lambda$ (e.g., Ratra & Quillen 1992).

We acknowledge the very helpful advice of D. Alvarez, D. Cottingham, P. de Bernardis, M. Dragovan, S. Hancock, M. Lim, S. Platt, L. Piccirillo, T. Readhead, J. Ruhl, G. Tucker, and especially T. Banday, M. Devlin, K. Ganga, K. Górski, G. Griffin, J. Gundersen, B. Netterfield, S. Tanaka, as well as L. Page. This work was supported in part by NSF grant PHY89-21378.

## FIGURE CAPTIONS

Fig. 1.– CMB anisotropy bandtemperature predictions for models 7 and 12 (lower values at $l \sim 200$). Continuous lines are what would be seen by a series of ideal, Kronecker window ($W_l \propto \delta_{l,l_e}$), experiments, for the models normalized to the central values of the DMR, ecliptic coordinates, normalization of BS. Open squares (at the appropriate $l_e$, and only for model 7) are the predictions for the windows of the table, with horizontal lines terminating at $l_{e^{-0.5}}$, and with vertical, correlated, $1\sigma$ error bars from the DMR normalization of BS. If a flat $\Lambda$ model turns out to be a good description of the universe, it is unlikely that near-future CMB observations will, by themselves, be able to pin down the values of $h$ and $\Omega_B$ with much accuracy.

Fig. 2.– CMB anisotropy bandtemperature observations (placed at the appropriate $l_e$) for all individual windows with data in the table (including those in the footnotes). Open squares are detections that are at least $2\sigma$ away from 0, and triangles are $2\sigma$ upper limits (they are placed at the $2\sigma$ upper limit, not at the peak of the likelihood). Vertical $1\sigma$ error bars also account for absolute calibration uncertainty. The continuous lines are the models of Fig. 1. As $l_e$ is not particularly physical, the observational data points should be directly compared to the prediction points of Fig. 1.

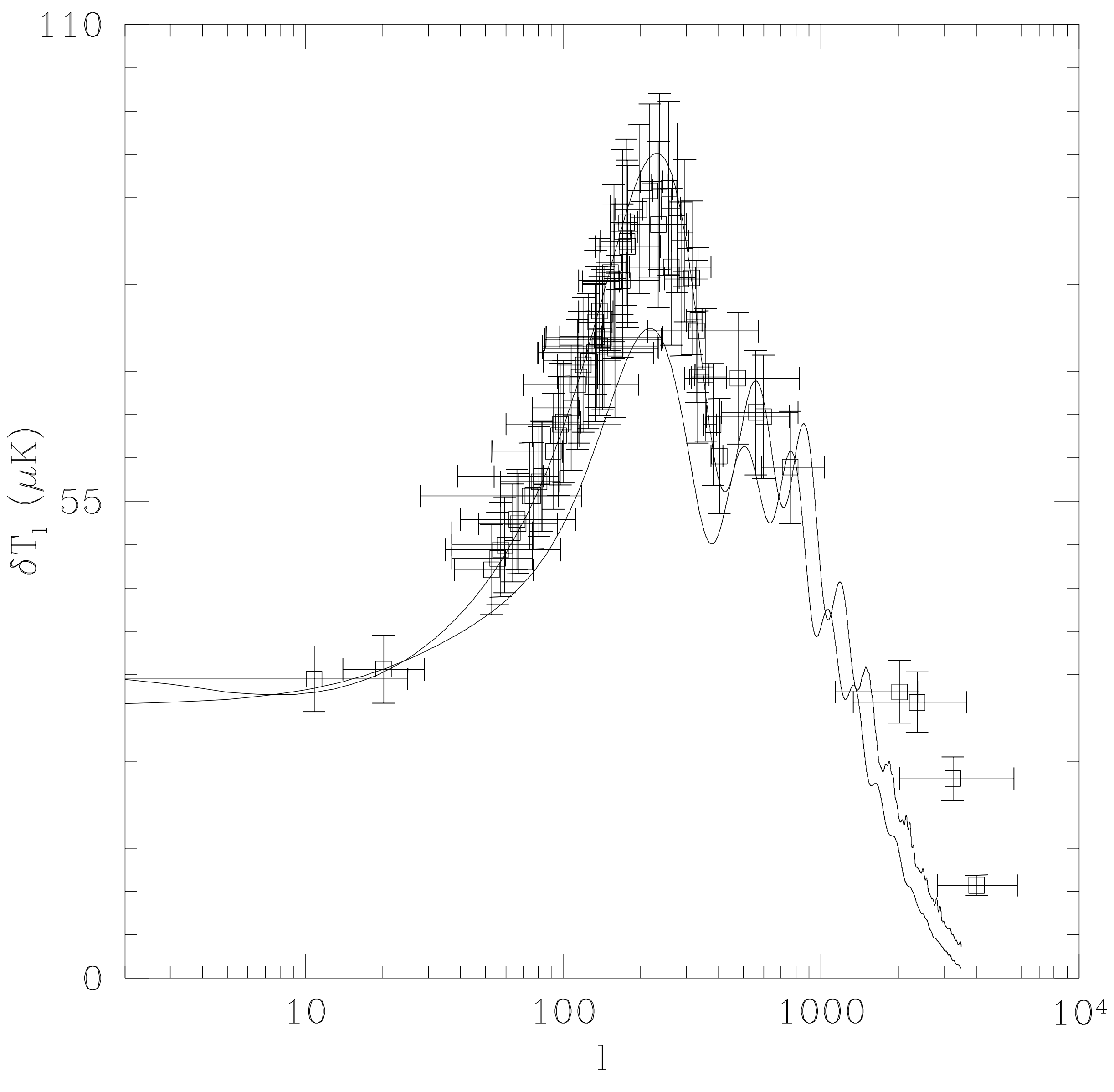

Figure 1

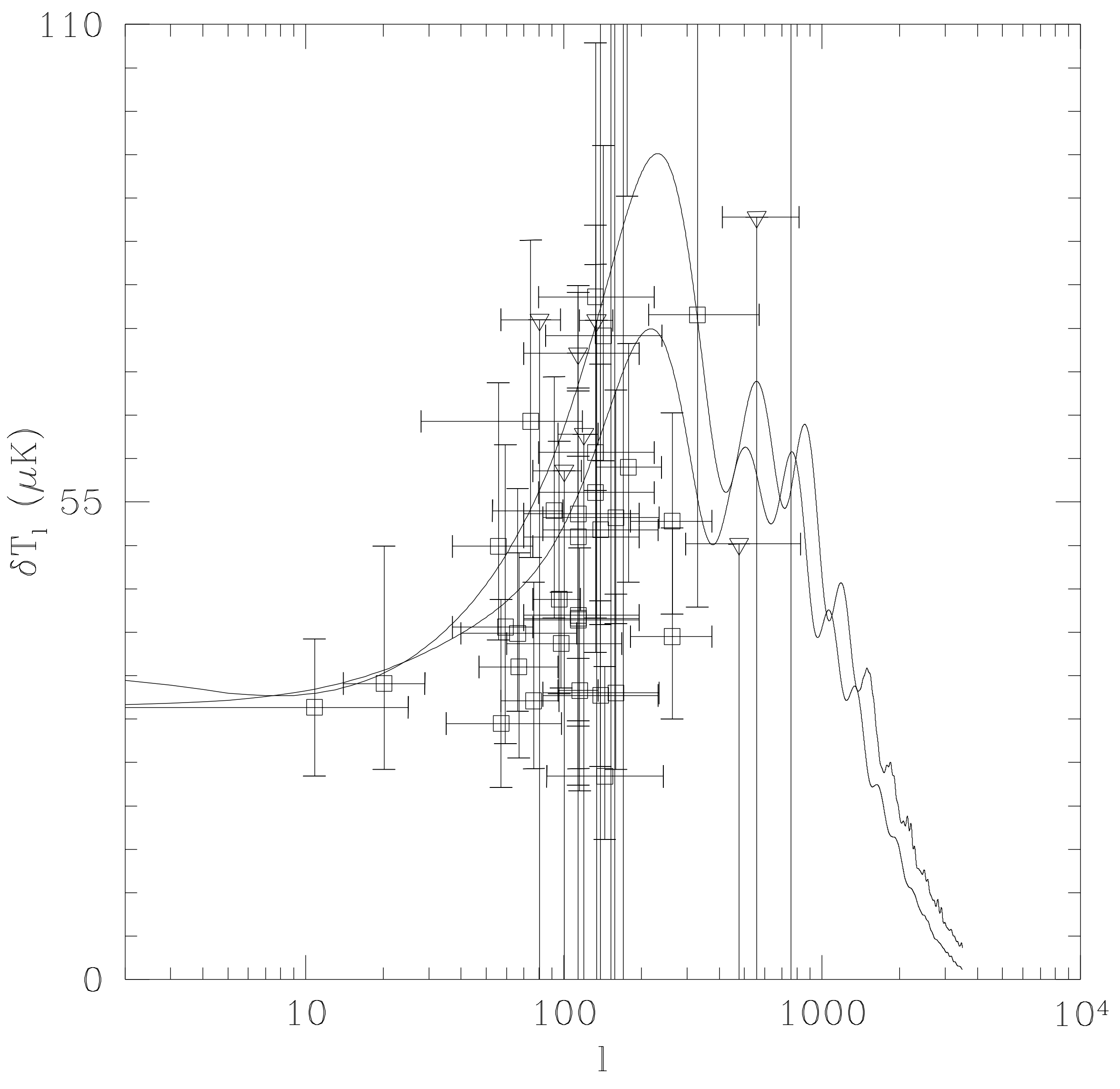

Figure 2